\documentclass[12pt]{article}

\def\rdots{\mathinner{\mkern1mu\raise1pt\vbox{\kern1pt\hbox{.}}\mkern2mu
   \raise4pt\hbox{.}\mkern2mu\raise7pt\hbox{.}\mkern1mu}}
\newcommand{\Z}{{\rm Z\kern-.35em Z}}
\newcommand{\bP}{{\rm I\kern-.15em P}}
\newcommand{\Q}{\kern.3em\rule{.07em}{.65em}\kern-.3em{\rm Q}}
\newcommand{\R}{{\rm I\kern-.15em R}}
\newcommand{\h}{{\rm I\kern-.15em H}}
\newcommand{\C}{\kern.3em\rule{.07em}{.65em}\kern-.3em{\rm C}}
\newcommand{\T}{{\rm T\kern-.35em T}}

\newcommand{\be}{\begin{equation}}
\newcommand{\ee}{\end{equation}}

\newcommand{\la}{\lambda}

\begin{document}

\openup 1.5\jot
\centerline{Dimer $\la_3 = .453 \pm .001$ and Some Other Very Intelligent Guesses}

\vspace{1in}
\centerline{Paul Federbush}
\centerline{Department of Mathematics}
\centerline{University of Michigan}
\centerline{Ann Arbor, MI 48109-1043}
\centerline{(pfed@umich.edu)}

\vspace{1in}

\centerline{\underline{Abstract}}

Working with a presumed asymptotic series for $\la_d$ developed in previous work, we make some intelligent guesses for $\la_d$ with $d=3, 4, 5$; and estimates for the corresponding errors.  We present arguments in favor of these guesses, we earnestly  believe they will turn out to be correct.  Such approximate values may help stimulate people working on rigorous bounds.  In addition to suggesting bounds to prove, there will be the strong motivation to prove me wrong.
\vfill\eject

In a previous paper, [1], we developed an asymptotic expansion for $\la_d$ of the dimer problem in powers of $1/d$.  In [2] computer calculations were performed to obtain some terms in this expansion, and also related quantities arising in the theory.  Using results in [2] we herein will argue for the following estimates.
\begin{eqnarray}
\la_2 &=& .296 \pm .007 \\
\la_3 &=& .453 \pm .001 \\
\la_4 &=& .5748 \pm .0006\\
\la_5 &=& .6785 \pm .0001 
\end{eqnarray}
This entire note is to argue for (2), (3), and (4).  The resut for $\la_2$ is exactly known, consistent with (1), which is included as the bellwether example of our algorithm to obtain (1) - (4).

In [2] we obtained in dimensions 2 and 3 a series $B_0, ..., B_5$, eq. (41)-(52) of [2].  Using eq. (22)-(27) of [2] and the theory developed in [1] one can obtain such a series $B_0,...,B_5$ in any dimension.  Our algorithm to obtain (1)-(4) above, in given dimension $d$, is to seek the two successive $B's$, $B_g$ and $B_{g+1}$, with minimum value of $|B_g - B_{g+1}|$.  Then with
\be	a = \frac 1 2 \big( B_g + B_{g+1} \big)   \ee
and
\be   b = |B_g - B_{g+1}|	\ee
our estimate is
\be   \la_d = a \pm b	. \ee

The following table encapsulates the $B$ series for $d=2,3,4,5$.
\[
\begin{tabular}{c|c|c|c|c|}
   & $d=2$ & $d=3$ & $d=4$ & $d=5$ \\ \hline
& & & &  \\
$B_0$ & .1931 & .3959 & .5397 & .6513 \\
 & & & & \\
$B_1$ & .2556 & .4375 & .5710 & .6763 \\
 & & & & \\ 
$B_2$ & .2921 & .4538 & .5801 & .6821 \\
 & & & & \\ 
$B_3$ & .2993 & .4524 & .5781 & .6803 \\
 & & & \\ 
$B_4$ & .2906 & .4468 & .5751 & .6786 \\
 & & & & \\ 
$B_5$ & .2814 & .4445 & .5745 & .6785
\end{tabular}   \]
It is generally believed that it is hard to compute the $\la_d, \ d > 2$, with great precision.  We note that the elements $B_5$ in our table required weeks of computer time to evaluate, and we question whether the $B_6$ terms will ever be computed.  We are only claiming $\la_3$, say, with a certain accuracy; we do not know how to get $\la_3$ with greater accuracy.  Because we do not have the $B_6$ terms, it might be better to double the error bounds of (3) and (4) above.

We give the following arguments in favor of our algorithm, (5)-(7) above.

1)  The $B_i$ term in the series of $B's$ is the sum of the first $i$ terms in a power series in $x$ (set equal to 1) that is presumed asymptotic.  Equations (5)-(7) encode the general rule of thumb wisdom for extracting a `best value' of the sum of an asymptotic expansion in a simple way.

2)  This algorithm gives a correct estimate for $\la_2$, eq. (1), which is known exactly.  We take this as rather compelling.

3)  $B_g$ as a function of $d$ is an expansion in powers $1/d$ up to power $1/d^g$.  The asymptotic expansion of [1] for $\la_d$ in powers of $1/d$ yields a `best approximation' in high dimensions that is a power series in $1/d$ to a high power.  For the approximation of our algorithm to match this approximation, $g$ {\it must increase with dimension}.

We have found this last argument hard to explain here, and it would also be hard to write down a precise statement corresponding to the discussion (though it could be done).  But for us the fact that $g=2$ for $d=2$ and $d=3$, and $g=4$ for $d=4$ and $d=5$, was as important as arguments 1) and 2) above.  It gives the algorithm a final ring of truth.

\vspace{1in}

\centerline{\underline{References}}

\begin{itemize}
\item[[1]] Paul Federbush, Hidden Structure in Tilings, Conjectured Asymptotic Expansion for $\la_d$ in Multidimensional Dimer Problem, \ arXiv : 0711.1092V9 [math-ph].
\item[[2]] Paul Federbush, Dimer $\la_d$ Expansion Computer Computations, \ arXiv : 0804.4220V1 [math-ph].

\end{itemize}

\end{document}